\documentclass[trackchanges,twocolumn]{aastex701}
\usepackage{CJK}

\newcommand{\msun}{{\rm M}_\odot}
\newcommand{\rsun}{{\rm R}_\odot}

\newcommand{\teffsun}{T_{\rm eff,\odot}}
\newcommand{\teffblt}{T_{\rm eff,bloated}}
\newcommand{\vcl}{\overline{v}}
\newcommand{\mave}{\overline{m}_*}
\newcommand{\dmcrit}{\dot{M}_{\rm crit}}

\newcommand{\fdir}{./}

\begin{document}
\begin{CJK*}{UTF8}{gbsn}

  \title{Mass and Spin Growth of Very Massive Stars in Star Clusters \\
    Potentially Associated with Little Red Dots}

  \author[orcid=0000-0002-8461-5517,sname='Tanikawa']{Ataru Tanikawa}
  \affiliation{Center for Information Science, Fukui Prefectural
    University, 4-1-1 Matsuoka Kenjojima, Eiheiji-cho, Fukui 910-1195,
    Japan}
  \email[show]{tanik@g.fpu.ac.jp}

  \author[orcid=0000-0002-4979-5671,sname='Shibata']{Masaru Shibata}
  \affiliation{Max-Planck-Institut f\"{u}r Gravitationsphysik
    (Albert-Einstein-Institut), Am M\"{u}hlenberg 1, D-14476
    Potsdam-Golm, Germany}
  \affiliation{Center for Gravitational Physics and Quantum
    Information, Yukawa Institute for Theoretical Physics, Kyoto
    University, Kyoto, 606-8502, Japan}
  \email{}

  \author[orcid=0000-0002-3517-1956,sname='Ioka']{Kunihito Ioka}
  \affiliation{Center for Gravitational Physics and Quantum
    Information, Yukawa Institute for Theoretical Physics, Kyoto
    University, Kyoto, 606-8502, Japan}
  \email{}



  \begin{abstract}

    Using gravitational $N$-body simulations, we investigate the
    evolution of mass and spin for very massive stars (VMSs) in dense
    star clusters, which could subsequently evolve into Little Red
    Dots (LRDs). Our results show that VMS masses can reach
    $10^3$--$10^4\,M_\odot$, depending on the initial conditions of
    the host clusters. Notably, the VMS mass increases by up to a
    factor of three when accounting for the bloated state at the
    Hayashi track induced by stellar collisions, provided that this
    state is maintained at accretion rates exceeding $3 \times
    10^{-2}\,M_\odot\,{\rm yr}^{-1}$. In all cases, the spin of the
    VMS, when normalized to the dimensionless black hole (BH) spin
    parameter, exceeds $10$, although the mass and spin of VMSs after
    the post-main sequence phase could be decreased by the stellar
    evolution process.  We nonetheless demonstrate that VMSs formed in
    dense star clusters can be highly spinning. Such a rapidly
    spinning VMS is expected to collapse into an intermediate-mass BH
    surrounded by a massive accretion disk. This BH-disk system could
    trigger powerful explosions and emit burst gravitational waves,
    similar to those observed in GW190521 and GW231123, for which the
    remnant BH masses are estimated to be $\gtrsim 100\,M_\odot$.

  \end{abstract}

\keywords{\uat{Star clusters}{1567} -- \uat{N-body simulations}{1083}
  -- \uat{Massive stars}{732} -- \uat{Gravitational wave
    sources}{677}}


\section{Introduction}
\label{sec:Introduction}

Very massive stars (VMSs) with initial masses of $\sim
10^2$--$10^4\,M_\odot$ are predicted to collapse into black holes
(BHs) triggered by electron-positron pair-instability
\citep{1971reas.book.....Z, 2001ApJ...550..372F,
  2002ApJ...567..532H}. One of the primary formation channels for VMSs
is runaway stellar collisions within dense star clusters
\citep{1970ApJ...162..791S, 1978MNRAS.185..847B, 1990ApJ...356..483Q,
  2004Natur.428..724P, 2006MNRAS.368..121F,
  2006MNRAS.368..141F,2015MNRAS.454.3150G, 2021MNRAS.501.5257R,
  2021MNRAS.507.5132D, 2022MNRAS.515.5106W, 2022ApJ...940..131G,
  2024Sci...384.1488F, 2025A&A...704A.321V}. In this scenario,
multiple stars undergo successive collisions to form a single VMS
before reaching the end of their stellar evolution. If such a VMS
collapses into a BH of comparable mass, the resulting object would be
classified as an intermediate-mass BH (IMBH). These IMBHs are
potential candidates for ultra-luminous and hyper-luminous X-ray
sources \citep[e.g.,][]{2001ApJ...547L..25M, 2009Natur.460...73F}, and
may explain the nature of the massive dark object recently identified
at the center of $\omega$ Centauri \citep{2024Natur.631..285H}.

Recently, the \textit{James Webb Space Telescope} (JWST) has
identified a novel population of red, massive, compact objects known
as ``Little Red Dots'' \citep[LRDs;][]{2023ApJ...954L...4K,
  2023ApJ...959...39H, 2025ApJ...991...37A,
  2025ApJ...978...92L}. These LRDs are characterized by broad emission
lines \citep{2024ApJ...963..129M, 2024ApJ...964...39G,
  2024ApJ...968...38K, 2025ApJ...986..126K, 2025ApJ...986..165T}, yet
they lack detectable X-ray \citep{2023ApJ...954L...4K,
  2024ApJ...974L..26Y, 2024MNRAS.535..853J, 2024ApJ...969L..18A,
  2025MNRAS.538.1921M, 2025ApJ...991...37A}, radio emission
\citep{2025ApJ...986..130G, 2025A&A...693L...2P, 2026A&A...706A.372M},
and variablities \citep{2025ApJ...983L..26T, 2025ApJ...985..119Z,
  2025ApJ...995...24K, 2025arXiv250919585S, 2025arXiv251116082B,
  2026arXiv260413000L} with a handful of exceptions
\citep{2025A&A...698A.227F, 2025MNRAS.544.3900J, 2025arXiv251205180Z,
  2025arXiv250316596N, 2026MNRAS.545f2117D}.  Two primary
interpretations for LRDs have emerged: (i) the ``BH envelope model,''
where massive BHs are embedded within dense gas envelopes
\citep{2025A&A...701A.168D, 2025arXiv250316596N, 2025MNRAS.544.3407K,
  2026Natur.649..574R}, and (ii) the ``star-only model,'' which
invokes extremely dense stellar systems \citep{2024RNAAS...8..182L,
  2024RNAAS...8..207G, 2024ApJ...977L..13B, 2025ApJ...991...37A}. Even
within the framework of the BH envelope model, dense stellar systems
may be necessary to explain the ultraviolet continua of LRDs
\citep{2026ApJ..1000...90I}, and may supply gas to a BH envelope
through tidal disruption events \citep{2026ApJ..1000L..21K}. In such
extreme environments, a sequence of processes -- from runaway stellar
collisions to VMS formation and the subsequent creation of massive BHs
-- is likely inevitable \citep{2025ApJ...994...40P,
  2025ApJ...995...44E}.

The collapse of a VMS, particularly one with high spin, may produce
detectable electromagnetic (EM) and gravitational-wave (GW)
transients. A rapidly spinning VMS is expected to form a BH surrounded
by a massive accretion disk immediately following pair instability
\citep{2006MNRAS.366.1424D}. Such a BH-disk system may launch
relativistic jets \citep{2001ApJ...550..372F, 2011ApJ...726..107S,
  2015ApJ...810...64M} and trigger explosions driven by viscous
angular momentum transport \citep{2019ApJ...870...98U}. Furthermore,
the massive accretion disk is susceptible to non-axisymmetric
instabilities, leading to the emission of burst GWs. Recent studies
\citep{2021PhRvD.103f3037S, 2026ApJ...996...57S} suggest that such VMS
collapses could explain enigmatic GW transients like GW190521
\citep{2020PhRvL.125j1102A} and GW231123
\citep{2025ApJ...993L..25A}. These events involve BH masses that fall
within the so-called pair-instability mass gap
\citep{2020ApJ...900L..13A, 2020ApJ...902L..36F, 2021ApJ...912L..31W,
  2022ApJ...924...39M, 2022ApJ...937..112F}, assuming that they
originate from genuine BH mergers.  Runaway collisions provide a
viable mechanism for forming such highly spinning VMSs, as the stellar
spin is significantly enhanced by the orbital angular momentum of the
colliding progenitors.

In this paper, we investigate runaway collisions and VMS formation
within dense star clusters, focusing on two primary objectives. First,
we track the evolution of the spin angular momentum of VMSs formed in
these clusters. This analysis is crucial to determine whether VMS
collapse can trigger the EM and GW transients proposed in recent
literature. Second, we incorporate the effects of stellar bloating on
VMS growth. Collision products are expected to remain bloated over the
Kelvin-Helmholtz (KH) timescale. If a VMS experiences successive
collisions within this timeframe, it maintains an expanded radius and
a low effective temperature of $\sim 5000$~K at the so-called Hayashi
track \citep{2012ApJ...756...93H, 2013ApJ...778..178H}, which has been
proposed by \cite{1978RvMP...50..437L}. Such a bloated VMS possesses a
significantly larger cross-section, facilitating further
collisions. To date, only a few studies have incorporated this
bloating effect, and it has never been accounted for in numerical
simulations of runaway collisions within dense star clusters as
potential progenitors that subsequently evolve into LRDs. Although
\cite{2009ApJ...695.1421F} considered the core-halo structure of VMSs,
their models assumed temperatures exceeding $2 \times 10^4$~K
\citep{1999PASJ...51..417I}, which remain considerably higher than the
values expected for a fully bloated state at the Hayashi
track. \cite{2023MNRAS.521.3553R} examined the bloated state of VMSs
within the context of Population III star clusters. Presumably because
of the Population III star cluster context, their assumed cluster mass
is $\sim 10^4 \msun$, which is smaller than ours ($10^5\text{--}10^6
\msun$).

The remainder of this paper is organized as follows. In Section
\ref{sec:Method}, we describe our numerical methodology and physical
assumptions. In Section \ref{sec:Results}, we present the results of
our $N$-body simulations. Section \ref{sec:Discussions} provides a
discussion on the implications for the EM and GW counterparts produced
by the collapse of highly spinning VMSs. Finally, we summarize our
conclusions in Section \ref{sec:Conclusions}. Throughout this paper,
$c$ and $G$ denote the speed of light and gravitational constant,
respectively.

\section{Method}
\label{sec:Method}

We perform $N$-body simulations to investigate four types of dense
star clusters with distinct initial conditions. The common
characteristics of these models are as follows. For the initial
phase-space density, we adopt the King model with a concentration
parameter of $W_0=12$ \citep{1966AJ.....71...64K}. We do not account
for initial mass segregation or fractal configurations. Clusters are
assumed to be isolated, meaning that they are not embedded within the
tidal field of a host galaxy. The initial binary fraction is set to
$0\%$. We adopt a cluster metallicity of $Z=2 \times 10^{-3}$
($0.1\,Z_\odot$), since LRDs may be metal-poor
\citep{2026arXiv260409177I}. These four cluster types vary in their
initial stellar mass functions (IMFs), total masses, and initial mass
densities within their half-mass radii, as summarized in Table
\ref{tab:initial_conditions}.  Regarding stellar populations, we
employ two types of IMFs. The first is the Kroupa IMF
\citep{2001MNRAS.322..231K}, with minimum and maximum stellar masses
of $0.08\,M_\odot$ and $150\,M_\odot$, respectively. The second is a
``top-heavy'' IMF defined as
\begin{equation}
  \frac{dN}{dm} \propto m^{-1} \quad (10 \le m/M_\odot \le 150),
\end{equation}
where $N$ and $m$ denote the stellar number and mass, respectively,
which is inspired by IMFs in low-metallicity and high-redshift
environments \citep{2021MNRAS.508.4175C, 2022MNRAS.514.4639C}.

The initial mass density within the half-mass radius of our cluster
models ranges from $10^6$ to $10^7\,M_\odot\,\text{pc}^{-3}$ (see the
column $\rho_{\rm h,i}$ in Table \ref{tab:initial_conditions}), which
are higher than those typically observed in the Milky Way and other
nearby galaxies. According to \cite{2010ARA&A..48..431P} and
\cite{2019ARA&A..57..227K}, the mass densities of such observed
clusters rarely exceed $10^6\,M_\odot\,\text{pc}^{-3}$, even in the
most extreme cases. Our cluster models are intended to represent the
central regions of the progenitor clusters of LRDs. Consequently, as
detailed below, while their mass densities are comparable to or even
higher than those of LRDs, their total masses remain smaller than the
LRD scale.

Within the framework of the star-only model, mass densities of
$10^4$--$10^5\,M_\odot\,{\rm pc}^{-3}$ within $1\,{\rm pc}$ are
typical values for LRDs, as derived by \cite{2024ApJ...977L..13B} and
\cite{2024RNAAS...8..207G}.  It is important to note that the number
of massive stars in the top-heavy IMF exceeds that of the Kroupa IMF
by a factor of 2 for stars with $>10\,M_\odot$ and by a factor of 15
for stars with $>100\,M_\odot$. The mass-to-light ratio of a cluster
with a top-heavy IMF is a factor of 10 smaller than that of a cluster
with a Kroupa IMF; for a given total mass, the former is 10 times
brighter than the latter. Although the actual mass density within
$1\,\text{pc}$ of the top-heavy cluster is $\sim
10^4\text{--}10^5\,M_\odot\,\text{pc}^{-3}$, it should be regarded as
equivalent to $\sim 10^5\text{--}10^6\,M_\odot\,\text{pc}^{-3}$ when
comparing it with the aforementioned literature. Notably, a
significant fraction of LRDs ($\sim 20\%$) may exhibit such high mass
densities \citep{2024RNAAS...8..207G}, although this statement relies
on a number of assumptions. In the context of a BH envelope model,
star clusters associated with LRDs are estimated to have $\sim
10^8$--$10^{10}\,M_\odot$ within $\lesssim 100\,{\rm pc}$
\citep{2025ApJ...986..126K}, corresponding to $\sim
20$--$2000\,M_\odot\,{\rm pc}^{-3}$.  While our cluster models appear
to have much higher mass densities than those typically associated
with LRDs in this model, it should be noted that the observed sizes of
LRD-associated clusters are upper limits; consequently, their derived
mass densities represent lower limits. Furthermore, since our cluster
models are significantly less massive than these observed systems, our
simulations effectively represent the dense central regions of such
clusters as progenitors of LRDs. This reasoning also applies to the
star-only model, where the total star cluster masses reach $\sim
10^8$--$10^9\,M_\odot$.

\begin{deluxetable}{cccccccccccc}
\tablewidth{0pt}
\tablecaption{Initial conditions and VMS results of $4$
  clusters.\label{tab:initial_conditions}}
\tablehead{
  \colhead{Model} & \colhead{$N_{\rm i}$} & \colhead{$M_{\rm i}$} & \colhead{$\rho_{\rm h,i}$} & \colhead{$\rho_{\rm 1pc,i}$} & \colhead{$t_{\rm rh,i}$} & \colhead{$t_{\rm dyn}$} & IMF & VMS bloated & VMS mass & VMS radius & VMS spin \\
  & & [$\msun$] & [$\msun\;{\rm pc}^{-3}$] & [$\msun\;{\rm pc}^{-3}$] & [year] & [year] & & & [$\msun$] & [$GM/c^2$] & [$GM^2/c$]
}
\startdata
M5D6K-N   & $1.7 \times 10^5$ & $10^5$ & $10^6$ & $2.4 \times 10^4$ & $1.1 \times 10^7$ & $4.3 \times 10^4$ & Kroupa    & No  & $1200$  & $1.7 \times 10^4$ & $28$ \\
M5D7K-N   & $1.7 \times 10^5$ & $10^5$ & $10^7$ & $2.4 \times 10^4$ & $3.4 \times 10^6$ & $1.3 \times 10^4$ & Kroupa    & No  & $2200$  & $1.6 \times 10^4$ & $10$ \\
M5D7H-N   & $2.0 \times 10^3$ & $10^5$ & $10^7$ & $2.4 \times 10^4$ & $6.7 \times 10^4$ & $2.2 \times 10^4$ & Top-heavy & No  & $4400$  & $1.4 \times 10^4$ & $10$ \\
M6D7H-N   & $1.9 \times 10^4$ & $10^6$ & $10^7$ & $2.4 \times 10^5$ & $4.8 \times 10^5$ & $1.6 \times 10^5$ & Top-heavy & No  & $24000$ & $4.8 \times 10^3$ & $4.0$ \\
M5D6K-Y   & $1.7 \times 10^5$ & $10^5$ & $10^6$ & $2.4 \times 10^4$ & $1.1 \times 10^7$ & $4.3 \times 10^4$ & Kroupa    & Yes & $5000$  & $2.0 \times 10^6$ & $56$ \\
M5D7K-Y   & $1.7 \times 10^5$ & $10^5$ & $10^7$ & $2.4 \times 10^4$ & $3.4 \times 10^6$ & $1.3 \times 10^4$ & Kroupa    & Yes & $8600$  & $1.7 \times 10^6$ & $14$ \\
M5D7H-Y   & $2.0 \times 10^3$ & $10^5$ & $10^7$ & $2.4 \times 10^4$ & $6.7 \times 10^4$ & $2.2 \times 10^4$ & Top-heavy & Yes & $19000$ & $1.5 \times 10^6$ & $18$ \\
M6D7H-Y   & $1.9 \times 10^4$ & $10^6$ & $10^7$ & $2.4 \times 10^5$ & $4.8 \times 10^5$ & $1.6 \times 10^5$ & Top-heavy & Yes & $95000$ & $1.2 \times 10^6$ & $6.5$ \\
\enddata

\tablecomments{$N_{\rm i}$, $M_{\rm i}$, $\rho_{\rm h,i}$, $\rho_{\rm
    1pc,i}$, and $t_{\rm rh,i}$ denotes the number of stars, cluster
  mass, mass density within half-mass radius, mass density within 1
  pc, and half-mass relaxation time at the initial time.}

\end{deluxetable}

We follow the dynamics of these clusters using the $N$-body simulation
code \texttt{PETAR} \citep{2020MNRAS.497..536W}. \texttt{PETAR} is
highly optimized for parallel computing via \texttt{FDPS}
\citep{2016PASJ...68...54I, 2020PASJ...72...13I} and accurately
resolves few-body orbits using the \texttt{SDAR} algorithm
\citep{2020MNRAS.493.3398W}. The code handles single and binary
stellar evolution through the \texttt{BSE} package
\citep{2000MNRAS.315..543H, 2002MNRAS.329..897H, 2020A&A...639A..41B},
which incorporates stellar wind mass loss based on the prescription of
\cite{2010ApJ...714.1217B}. In our simulations, two stellar wind
regimes are particularly prominent: the winds of hot, massive H-rich
stars \citep{Vink01} and those of luminous blue variables
\citep[LBVs;][]{1994PASP..106.1025H}. In particular, the mass loss
rate of the LBV wind is given by
  \begin{eqnarray}
    \dot{M}_{\rm LBV} = 1.5 \times 10^{-4}\;\msun\;{\rm yr}^{-1},
  \end{eqnarray}
when $L > 6 \times 10^5$ and $10^{-5} R L^{1/2} > 1$ are satisfied,
where $L$ and $R$ are stellar luminosity and radius in solar units,
respectively. This mass-loss rate prescription was originally
developed for stars with $\lesssim 100\,\msun$, and applying it to
VMSs may underestimate the actual rates. Therefore, the wind mass loss
in our models should be interpreted as a conservative lower
limit. Furthermore, unlike the analytical approach taken by
\cite{2025ApJ...994...40P}, we do not implement specialized mass-loss
treatments for VMSs. We assume perfectly sticky collisions, neglecting
any loss of mass or angular momentum specifically driven by the
stellar collision process itself.

We employ a single-star evolution model that differs from the default
prescription in the \texttt{BSE} code, which we refer to as the ``L
model.'' This model was constructed based on 1D stellar evolution
simulations of stars with masses ranging from $8\,M_\odot$ to
$1280\,M_\odot$, performed using the \texttt{HOSHI} code
\citep{2016MNRAS.456.1320T, 2018ApJ...857..111T, Takahashi19,
  Yoshida19}. The detailed parameter settings and physical assumptions
of the L model are described in \cite{2020MNRAS.495.4170T} and
\cite{2022ApJ...926...83T}.

We follow the cluster evolution for $1\,{\rm Myr}$ after formation,
ensuring that all stars remain in the main-sequence (MS) phase
throughout the simulations. In our model, a collision between two MS
stars is triggered when their separation becomes smaller than the sum
of their radii. Upon merging, the collision product undergoes
rejuvenation due to the supply of fresh hydrogen to its core. We
quantify this rejuvenation process as
\begin{eqnarray}
  T_{\rm age,3} = C T_{\rm ms,3} \frac{m_1(T_{\rm age,1}/T_{\rm ms,1})
    + m_2(T_{\rm age,2}/T_{\rm ms,2})}{m_3}.
\end{eqnarray}
Here, $m_i$ represents the mass of the $i$-th star, and the indices
$1$ and $2$ denote the progenitor stars ($m_1 \ge m_2$), while the
index $3$ denotes the resulting collision product (i.e.,
$m_3=m_1+m_2$). $T_{\mathrm{ms},i}$ is the main-sequence lifetime of a
stellar model with an initial mass $m_i$ calculated without mass
loss. $T_{\mathrm{age},i}$ is the current age of the $i$-th star; in
particular, $T_{\mathrm{age},3}$ represents the effective age of a
stellar model if it were to evolve with a constant mass $m_3$.  The
rejuvenation coefficient $C$ is defined as
\begin{eqnarray}
  C = \left\{
  \begin{array}{ll}
    0.1 & (m_2/m_1 > 0.1) \\
    1.0 & (m_2/m_1 \le 0.1)
  \end{array}
  \right..
\end{eqnarray}
Under this prescription, the collision product is significantly
rejuvenated if the progenitor stars have comparable masses, whereas
rejuvenation is suppressed for highly unequal-mass collisions. For
instance, this prevents unrealistic rejuvenation in collisions between
a $1000\,M_\odot$ VMS and a $1\,M_\odot$ star. Note that the original
\texttt{BSE} code adopts a constant $C=0.1$ regardless of the mass
ratio $m_2/m_1$, which tends to over-rejuvenate products in encounters
with very small ratio, $m_2/m_1$.

The rejuvenation coefficient $C=0.1$ potentially causes
over-rejuvenation. However, its impact remains limited due to the
following two reasons. First, while the lifetime of an initially VMS
would exceed $1\,\mathrm{Myr}$, our simulations only track the
evolution up to $1\,\mathrm{Myr}$. Therefore, even if
over-rejuvenation occurs, it does not affect the current
results. Second, the majority of mergers involve mass ratios of
$m_2/m_1 \le 0.1$; in such cases, $C=1$ is selected instead of
$C=0.1$. If the simulation timescale is extended in future work, this
coefficient warrants further investigation.

\begin{figure}[ht!]
  \plotone{\fdir/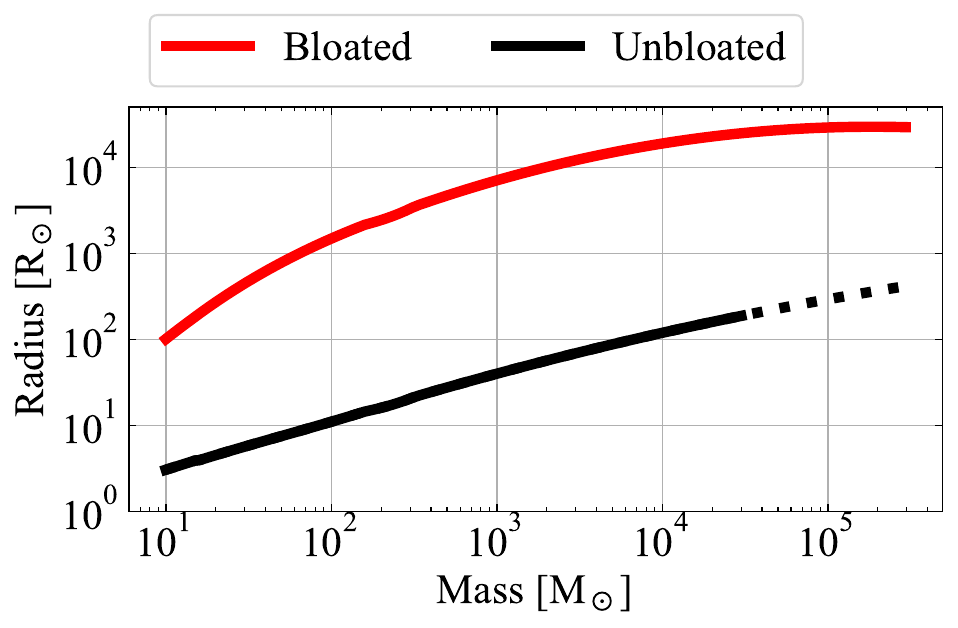}
  \caption{Relation between stellar mass and radius at unbloated and
    bloated states at the ZAMS time. The dotted curve denotes
    unbloated radii for $> 30000\;\msun$. These data points are not
    utilized, as the VMSs are too compact to be stable to general
    relativistic instability with mass $\gtrsim30000\;\msun$ for the
    unbloated cases. \label{fig:bloatedradius}}
\end{figure}

Stellar collisions convert kinetic energy into thermal energy within
the resulting collision product. Consequently, the VMS shifts onto the
Hayashi track and remains in a bloated state throughout its KH
timescale.  If a VMS successively accretes other stars within this KH
timeframe, it sustains this expanded state \citep{2012ApJ...756...93H,
  2013ApJ...778..178H}, which in turn increases the cross-section for
further stellar collisions.

We model the bloated state of a VMS as follows. Since the maximum
initial stellar mass in our clusters is $150\,M_\odot$, any star
exceeding $M > 150\,M_\odot$ is considered a collision product. For
these objects, we prescribe a persistently bloated state, where the
stellar radius is determined by fixing the effective temperature at
$\teffblt = 5000$~K. In this regime, we assume that the luminosity is
preserved from the unbloated state ($L_{\rm bloated} = L_{\rm
  unbloated}$), where $L_{\rm unbloated}$ is derived from our stellar
evolution model without considering the bloated state.  Specifically,
the bloated radius $R_{\rm bloated}$ is expressed as
\begin{eqnarray}
  \left( \frac{R_{\rm bloated}}{R_\odot} \right) = 4.2 \times 10^{3}
  \left( \frac{L_{\rm bloated}}{10^{7.5}\,L_\odot} \right)^{1/2}
  \left( \frac{\teffblt}{\teffsun} \right)^{-2},
\end{eqnarray}
where $\teffsun=5777$~K. As a VMS evolves into a post-MS star, it may
expand beyond this radius. Therefore, the actual VMS radius $R_{\rm
  VMS}$ is given by
\begin{eqnarray}
  R_{\rm VMS} = \max(R_{\rm bloated}, R_{\rm unbloated}),
\end{eqnarray}
where $R_{\rm unbloated}$ represents the radius in the unbloated
case\footnote{This is because the effective temperature of a star can
fall below 5000 K in our single-star evolution model. However, it does
not drop below 5000 K in the current simulations, and $R_{\rm bloated}
> R_{\rm unbloated}$ always holds true.}.  We maintain this bloated
state even if the VMS does not undergo a collision strictly within one
KH timescale; however, we subsequently perform a post-processing
analysis to assess the self-consistency of this assumption.

In our post-processing analysis, we evaluate the self-consistency of
the bloated state model using three distinct criteria. The first
criterion is whether the VMS sustains an accretion rate exceeding
$\dmcrit = 3 \times 10^{-2}\,M_\odot\,{\rm yr}^{-1}$
\citep{2012ApJ...756...93H, 2013ApJ...778..178H}.  The remaining two
criteria pertain to the KH timescale of the VMS, for which we consider
two different physical cases. Specifically, we calculate the KH
timescale by adopting either the bloated radius ($R_{\rm bloated}$) or
the unbloated radius ($R_{\rm unbloated}$), while maintaining the
unbloated luminosity ($L_{\rm unbloated}$) consistently in both
cases. These KH timescales are given by
\begin{eqnarray}
  \left( \frac{t_{\rm KH, unbloated}}{\rm yr} \right) &=& 2.5 \times
  10^4 \left( \frac{M}{10^3\,M_\odot} \right)^2 \nonumber \\
  & & \times \left( \frac{R_{\rm unbloated}}{10^{1.6}\,R_\odot}
  \right)^{-1} \left( \frac{L_{\rm unbloated}}{10^{7.5}\,L_\odot}
  \right)^{-1}, \label{eq:tkhunbloated} \\
  \left( \frac{t_{\rm KH, bloated}}{\rm yr} \right) &=& 2.0 \times
  10^2 \left( \frac{M}{10^3\,M_\odot} \right)^2 \nonumber \\
  & & \times \left( \frac{R_{\rm bloated}}{10^{3.7}\,R_\odot}
  \right)^{-1} \left( \frac{L_{\rm unbloated}}{10^{7.5}\,L_\odot}
  \right)^{-1}. \label{eq:tkhbloated}
\end{eqnarray}
The use of $t_{\rm KH, unbloated}$ as a criterion may be physically
justified because the bulk of the VMS mass remains centrally
concentrated, even when the envelope is bloated
\citep{2012ApJ...756...93H, 2013ApJ...778..178H}.

The collision criterion for a bloated VMS is identical to that used
for standard MS-MS collisions: a collision is triggered when the
separation between two stars becomes smaller than the sum of their
radii. We assume that these collisions are perfectly sticky, resulting
in no mass loss, and that the orbital angular momentum of the
progenitor binary is entirely partitioned into the spin angular
momentum of the collision product. It should be noted, however, that
since a VMS likely possesses a distinct core-envelope structure, a
stellar collision could potentially strip the envelope. In such a
scenario, the orbital angular momentum might not be fully converted
into the spin of the resulting VMS. By adopting the current assumption
of full conversion and zero mass loss, our model effectively provides
an upper limit for both the mass and spin growth of the VMS.

Figure \ref{fig:bloatedradius} illustrates the relationship between
stellar mass and radius for zero-age main-sequence (ZAMS) stars in
both unbloated and bloated states. As shown, the bloated radii
approach a constant value for masses $\gtrsim 10^5\,\msun$. This
unreasonable thing happens for the following reason. Our single-star
evolution model is based on fitting formulae derived from 1D
simulation results for stars ranging from $8\,\msun$ to
$1280\,\msun$. While these formulae are extrapolated to the extremely
high-mass regime, it should be noted that this extrapolation may no
longer be physically robust for stars exceeding $\gtrsim 10^5\,\msun$.

For each initial condition, we conduct two separate simulations: one
applying the bloated state to collision products and another omitting
it. These cases are labeled as ``Yes'' or ``No'' in the ``VMS
bloated'' column of Table \ref{tab:initial_conditions}, and are
further distinguished by the suffixes ``Y'' and ``N'' in their
respective model names. In total, we perform eight simulation runs.

\section{Results}
\label{sec:Results}

Runaway collisions occur in all our models, and VMS masses exceed
$1000\;\msun$, as the dynamical friction timescale for the most
massive stars is sufficiently shorter than their or VMS evolutionary
timescales\footnote{If this hierarchy of timescales is reversed,
runaway collisions become unlikely. In such cases, massive stars
evolve into BHs before they can sink to the cluster center. Once
formed, BHs have a collisional cross-section that is orders of
magnitude smaller than that of stars, making further collisions with
other stars highly improbable. On the other hand, it has also been
pointed out that IMBHs could potentially form through pathways
distinct from runaway collisions
\citep[e.g.][]{2015MNRAS.454.3150G}.}. The dynamical friction
timescale can be expressed as
\begin{eqnarray}
  t_{\rm dyn} \sim \frac{m_{\rm ave,i}}{m_{\rm heavy,i}} t_{\rm rh,i},
\end{eqnarray}
where $m_{\rm heavy,i}$ and $m_{\rm ave,i}$ are the heaviest stellar
mass and average mass at the initial time, respectively, and $t_{\rm
  rh,i}$ is the initial half-mass relaxation time. This initial
relaxation time is defined as
\begin{eqnarray}
  t_{\rm rh,i} = 0.0477 \frac{N_{\rm i}}{(G \rho_{\rm h,i})^{1/2}
    \ln(0.4N_{\rm i})},
\end{eqnarray}
where $N_{\rm i}$ and $\rho_{\rm h,i}$ denote the initial number of
stars and the initial half-mass density, respectively. The dynamical
friction timescale for models M6D7H can be written as
\begin{eqnarray}
  t_{\rm dyn} \sim &1.6& \times 10^5 \left( \frac{m_{\rm
      ave}}{0.59\;\msun} \right) \nonumber \\
  &\times& \left( \frac{m_{\rm heavy}}{150\;\msun} \right)^{-1} \left(
  \frac{t_{\rm rh,i}}{11\;{\rm Myr}} \right)\;[{\rm yr}].
\end{eqnarray}
Note that models M6D7H have the longest dynamical friction timescale
among our models (see Table \ref{tab:initial_conditions}). The
dynamical friction timescale is much shorter than the VMS evolution
timescale ($\sim 3$ Myr). In general, the core relaxation time is used
as the criterion for the onset of runaway collisions. In this study,
however, we adopt the half-mass relaxation time because the stars
contributing to the runaway process originated largely from outside
the initial core. This is evident from the fact that the total mass of
stars initially residing within the core accounted for only $\sim
0.5\%$ of the cluster mass, whereas the final VMS mass reaches $\sim
1\text{--}10\%$ of the total mass.

\begin{figure*}[ht!]
  \plottwo{\fdir/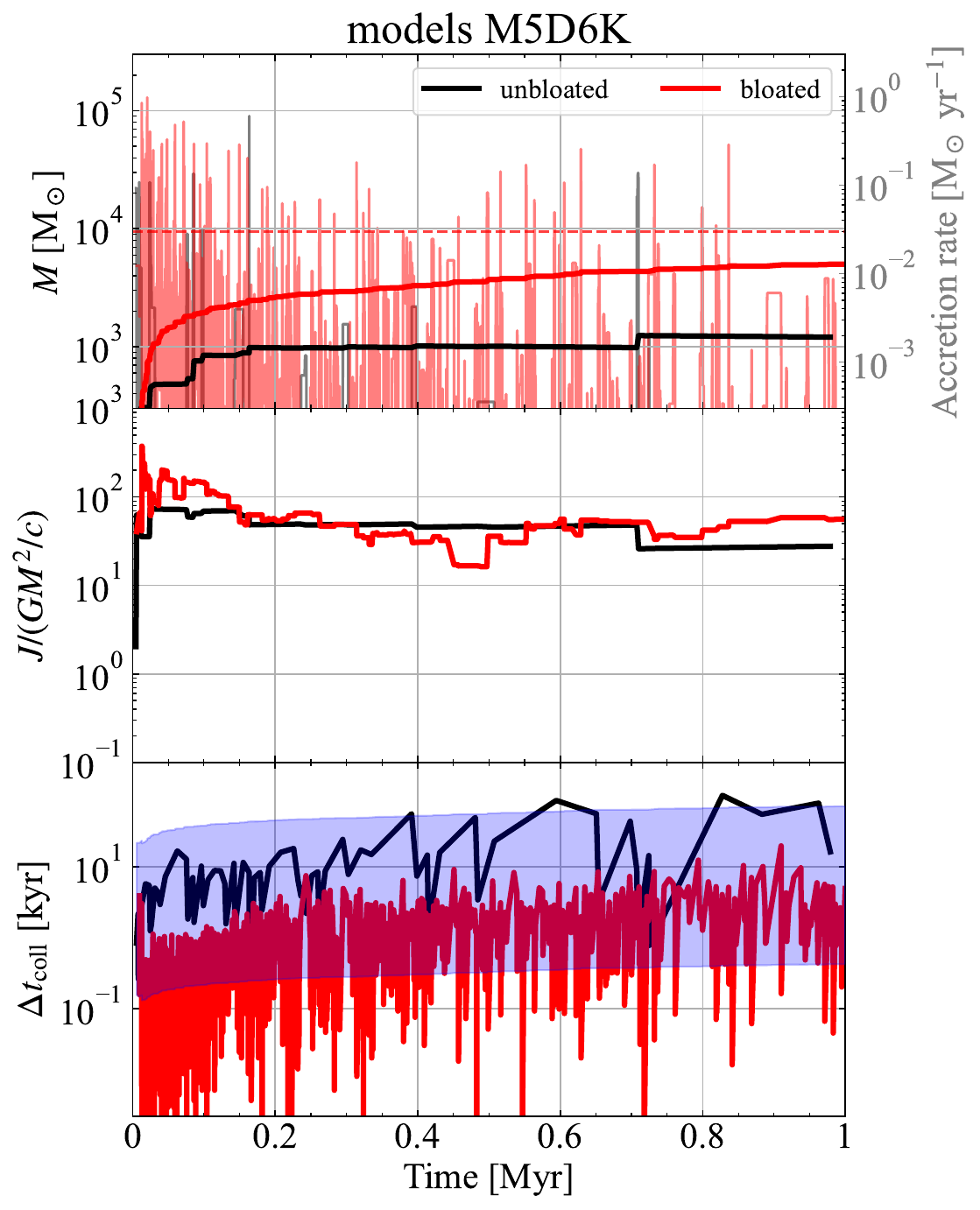}{\fdir/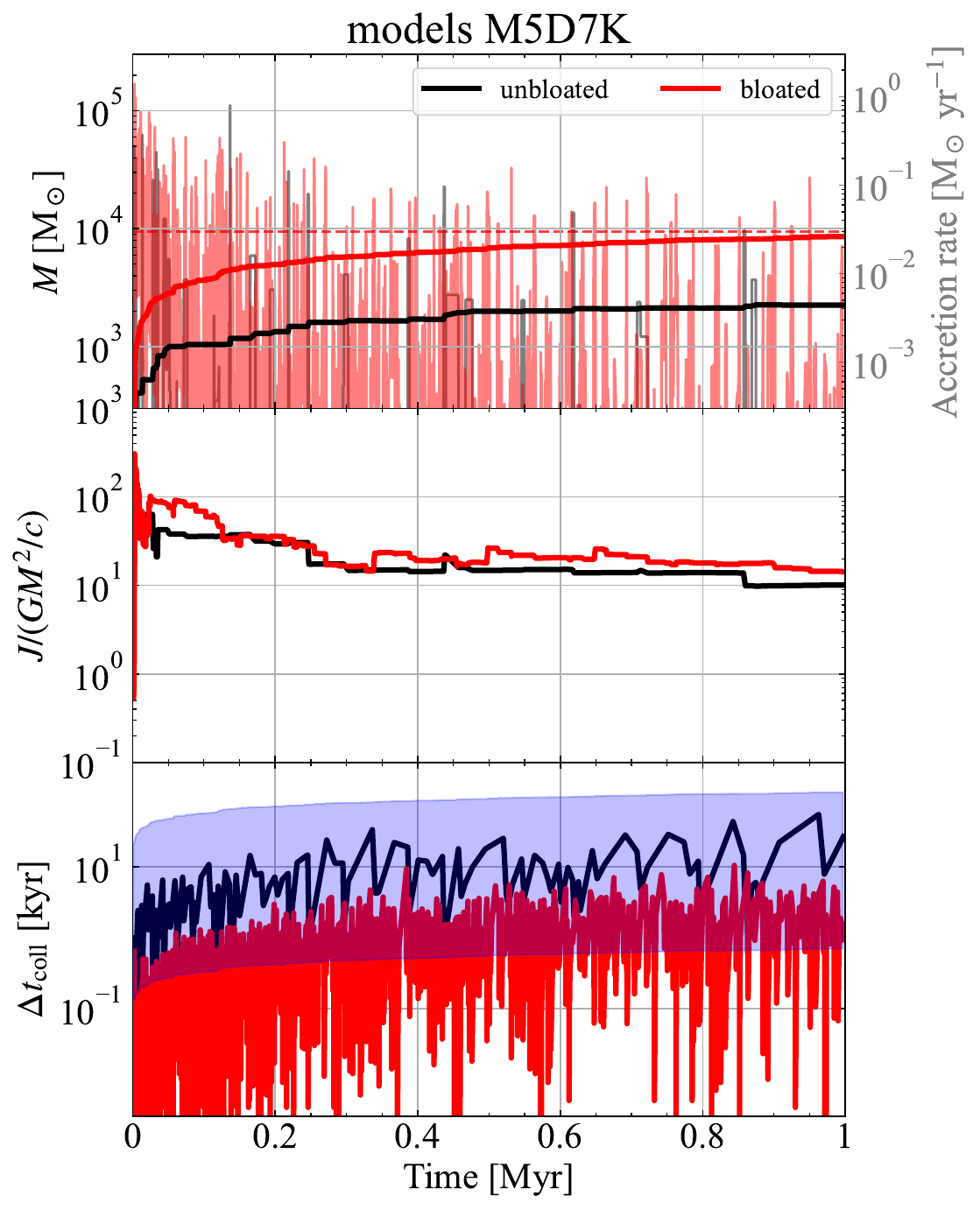}
  \plottwo{\fdir/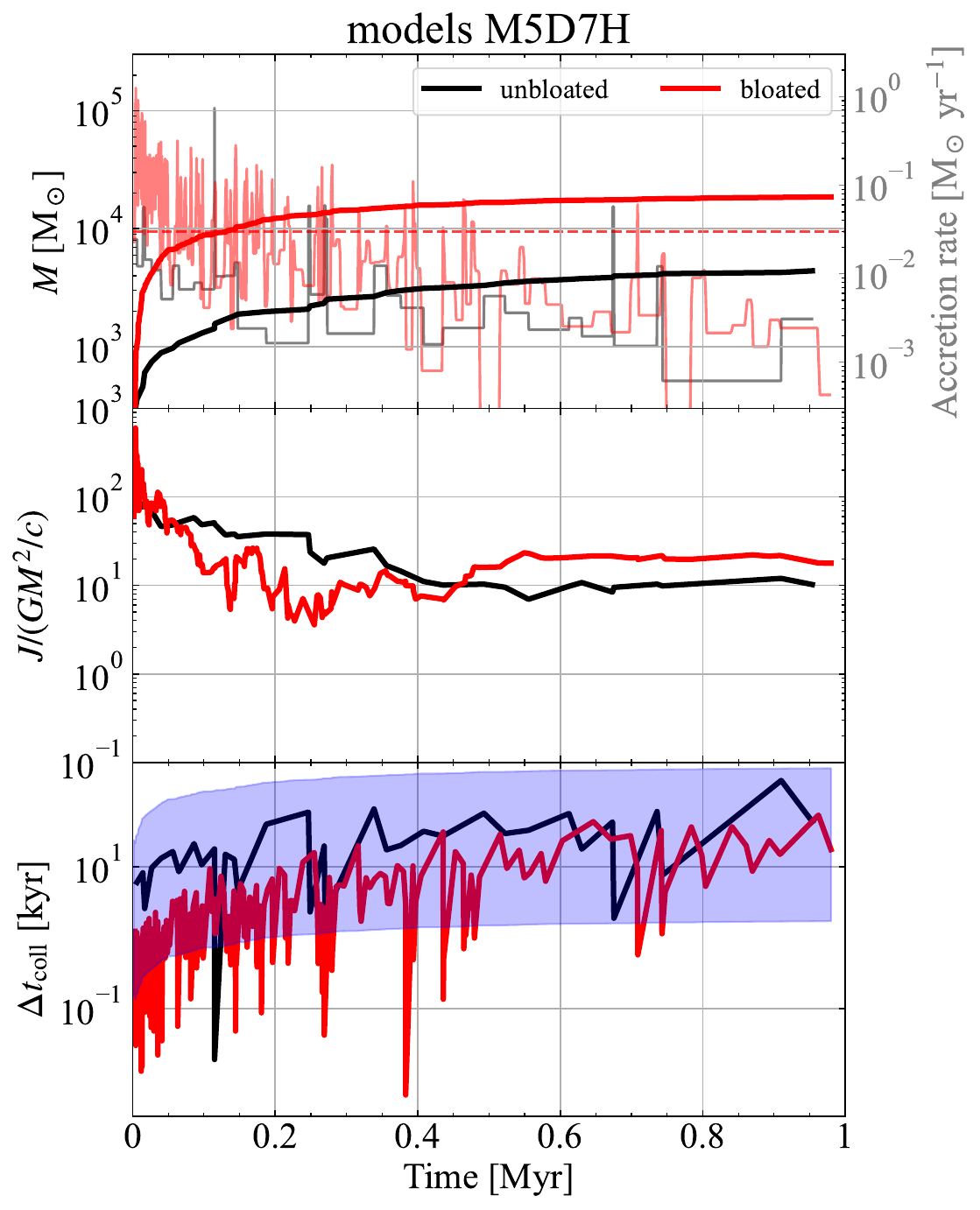}{\fdir/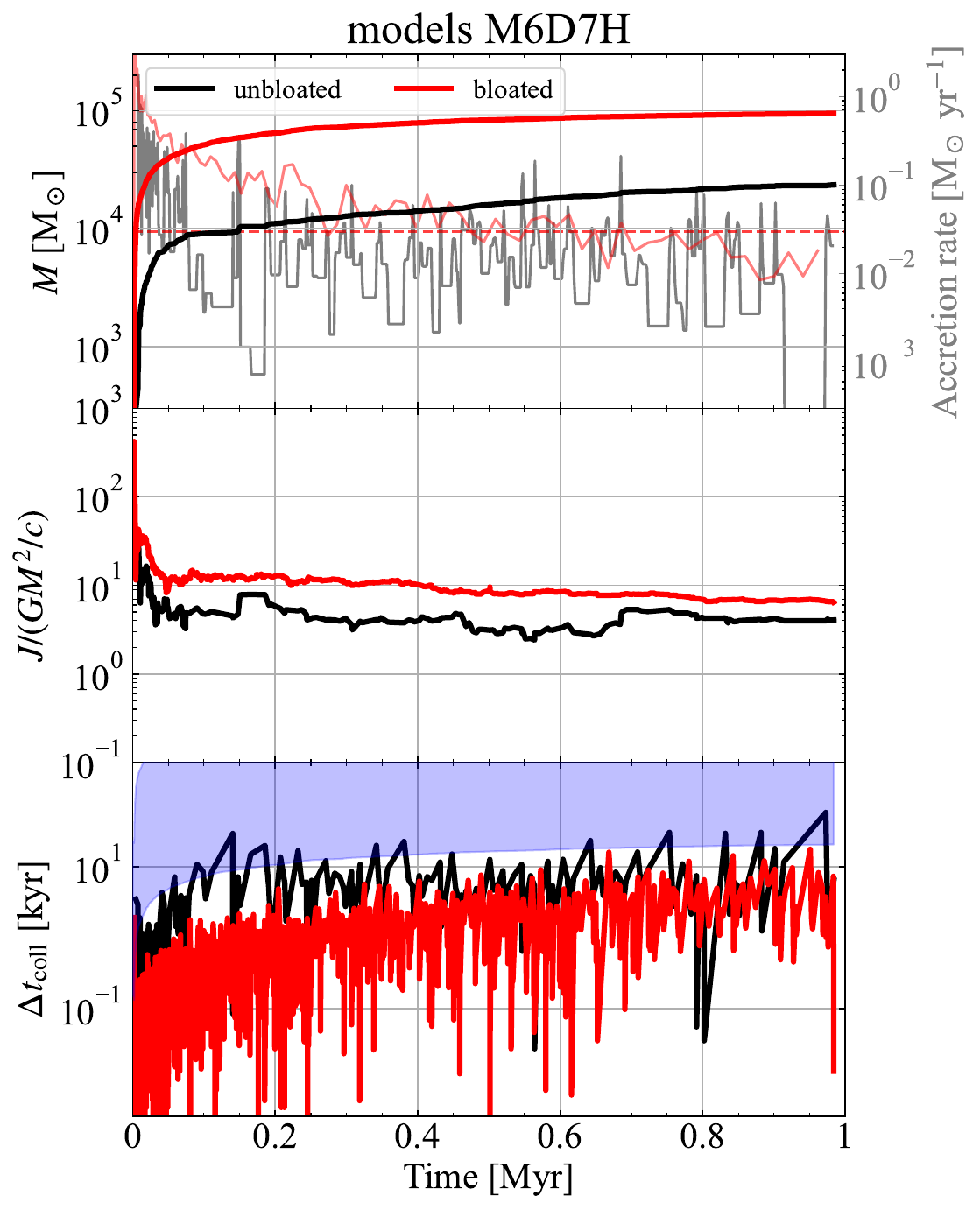}
  \caption{Time evolution of VMS masses, accretion rates,
    dimensionless spins, and time intervals between collisions for
    each model. Black and red curves indicate unbloated and bloated
    models, respectively. Thin grey and red curves in the top panels
    indicate VMS mass accretion rates in the unbloated and bloated
    cases, respectively. The shaded regions in the bottom panels show
    regions sandwiched between the minimum and maximum of KH
    timescales, i.e.  $t_{\rm KH,bloated}$ and $t_{\rm KH,unbloated}$,
    respectively (see Eq. (\ref{eq:tkhbloated}) and
    Eq. (\ref{eq:tkhunbloated})). The dashed lines indicate an
    accretion rate of $\dmcrit = 3 \times 10^{-2}\;\msun\;{\rm
      yr}^{-1}$, which is the lower limit for keeping the bloated
    states suggested by \cite{2012ApJ...756...93H,
      2013ApJ...778..178H}. The accretion rate is plotted at intervals
    equal to the instantaneous KH timescale. Checking the accretion
    rate at these intervals is necessary because the VMS returns to an
    unbloated state if the rate drops below the critical accretion
    rate during KH timescale \citep{2015MNRAS.452..755S}.
    \label{fig:mergerhistory}}
\end{figure*}

Figure~\ref{fig:mergerhistory} illustrates the mass evolution of the
most massive star (i.e., VMS) within each star cluster. In model
M5D7K-N, the VMS mass reaches $2.2 \times 10^3\,\msun$, accounting for
approximately $2.2$\% of the total cluster mass. The VMS grows rapidly
during the initial $0.1\,{\rm Myr}$, after which the growth rate
declines as the cluster expands and becomes more diffuse due to
two-body relaxation. By $1\,{\rm Myr}$, the mass growth has
effectively ceased. In contrast, for model M5D7K-Y, which incorporates
the bloated state of the VMS, the growth is significantly
enhanced. Due to its increased cross-section, the bloated VMS
undergoes mergers more frequently, reaching $\sim 8.6 \times
10^3\,\msun$ (or $8.6$\% of the cluster mass) by $1\,{\rm
  Myr}$. Beyond this point, the VMS is expected to grow further. The
physical validity of adopting this bloated state model is discussed in
detail later.

When considering either the bloated or unbloated VMS cases
individually, Table~\ref{tab:initial_conditions} reveals that the mass
ratio of the VMS to the total cluster mass is consistently
anti-correlated with the initial half-mass relaxation time ($t_{\rm
  rh,i}$). In both scenarios, the mass ratio increases as the initial
mass density increases (cf. models M5D6K and M5D7K), as the IMF
becomes more top-heavy (cf. models M5D7K and M5D7H), or as the total
number of stars decreases (cf. models M5D7H and M6D7H). This
consistent trend arises because the runaway collision process is
fundamentally driven by two-body relaxation.

\begin{figure}[ht!]
  \plotone{\fdir/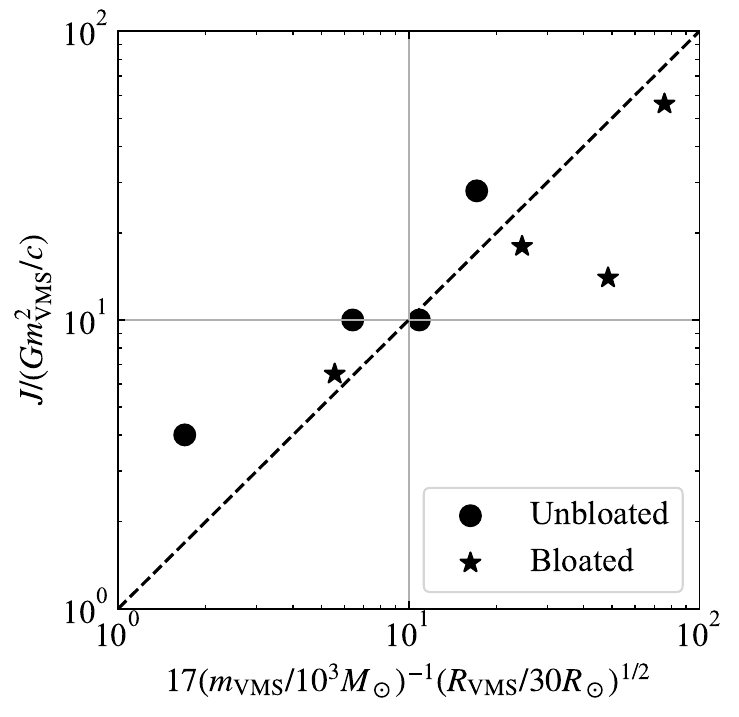}
  \caption{Relation between VMS spins in simulation results, and
    estimated with Eq. (\ref{eq:spinparam}). These values are equal on
    the dashed line. Circles and star marks indicate the unbloated and
    bloated cases, respectively. We assume
    $m_{*}=10\;\msun$. \label{fig:spinparam}}
\end{figure}

The spin angular momenta of the VMSs, normalized to the dimensionless
spin parameter, far exceed unity (see
Table~\ref{tab:initial_conditions} and
Figure~\ref{fig:mergerhistory}). These VMS spins can be estimated as
follows. The spin angular momentum of a VMS evolves through a random
walk, where each step corresponds to the orbital angular momentum
contributed by a collided star. The total angular momentum $J$ can be
approximated as
\begin{eqnarray}
  J \sim N_{\rm coll}^{1/2} \mave b \vcl, \label{eq:angularmomentum0}
\end{eqnarray}
where $N_{\rm coll}$ is the number of stellar collisions, $\mave$ is
the average mass of a star colliding with the VMS, $b$ is the typical
impact parameter, and $\vcl$ is the 3D velocity dispersion of the
cluster. The impact parameter $b$ can be estimated from the Keplerian
limit as
\begin{eqnarray}
  b \sim \frac{\sqrt{2Gm_{\rm VMS}R_{\rm VMS}}}{\vcl},
    \label{eq:impactparameter}
\end{eqnarray}
where $m_{\rm VMS}$ and $R_{\rm VMS}$ are the mass and radius of the
VMS, respectively. Substituting Eq.~(\ref{eq:impactparameter}) into
Eq.~(\ref{eq:angularmomentum0}), we obtain
\begin{eqnarray}
  J \sim N_{\rm coll}^{1/2} \mave \sqrt{2Gm_{\rm VMS}R_{\rm VMS}}.
    \label{eq:angularmomentum1}
\end{eqnarray}
Assuming $m_{\rm VMS} \sim N_{\rm coll}\mave$, we can rewrite this
expression as
\begin{eqnarray}
  J \sim m_{\rm VMS} \sqrt{2G \mave R_{\rm VMS}}.
    \label{eq:angularmomentum2}
\end{eqnarray}
Normalizing this result to the dimensionless spin parameter, we find
\begin{eqnarray}
  \frac{J}{Gm_{\rm VMS}^2/c} \sim 17 \left( \frac{\mave}{10\,\msun}
  \right)^{1/2} \left( \frac{m_{\rm VMS}}{10^3\,\msun} \right)^{-1}
  \left( \frac{R_{\rm VMS}}{30\,\rsun}
  \right)^{1/2}, \label{eq:spinparam}
\end{eqnarray}
where we have adopted the parameters characteristic of model M5D7K-N.
In Figure~\ref{fig:spinparam}, we compare the VMS spins obtained from
our simulations with the values estimated using
Eq.~(\ref{eq:spinparam}). We find that they are consistent within a
factor of a few.

In the following, we examine the simulation results for the cases
incorporating the bloated state, with a primary focus on model
M6D7H-Y. In this model, the accretion rate of the VMS exceeds $\dmcrit
= 3 \times 10^{-2}\,\msun\,{\rm yr}^{-1}$ during the initial
$0.5\,{\rm Myr}$. Meanwhile, the time intervals between successive
stellar collisions remain shorter than $t_{\rm KH, bloated}$
throughout the first $1\,{\rm Myr}$. Our analysis indicates that the
criterion that the VMS accretion rate exceeds $\dmcrit$ is the most
stringent.  If we strictly apply this condition, the bloated state
would be maintained only during the initial $0.5\,{\rm Myr}$. However,
even if the VMS follows our simulation results for the first
$0.5\,{\rm Myr}$ and subsequently ceases its rapid growth by
transitioning to an unbloated state, it would still reach a mass of at
least $8.4 \times 10^4\,\msun$. This corresponds to approximately
$10$\% of the total cluster mass, which is more than three times
larger than the mass attained in a scenario where the VMS remains
unbloated throughout its entire evolution.

For the three remaining models other than M6D7H-Y, the time intervals
between successive collisions are shorter than $t_{\rm KH, unbloated}$
but longer than $t_{\rm KH, bloated}$ (see
Figure~\ref{fig:mergerhistory}). Furthermore, for models M5D6K-Y and
M5D7K-Y, the accretion rate of the VMS only intermittently exceeds
$\dmcrit = 3 \times 10^{-2}\,\msun\,{\rm yr}^{-1}$. This potentially
implies that the bloated state is short-lived and of minor importance
in these models. For model M5D7H-Y, the accretion rate exceeds
$\dmcrit = 3 \times 10^{-2}\,\msun\,{\rm yr}^{-1}$ for most of the
time, albeit only during the first $0.05\,\mathrm{Myr}$. Within this
short period, the VMS mass has already reached twice that of the VMS
in model M5D7H-N at $1\,\mathrm{Myr}$.

Finally, we compare our results with several previous simulations of
extremely dense star clusters resembling LRDs.
\cite{2025ApJ...994...40P} reported the formation of $\sim
10^4\,\msun$ VMSs within star clusters of $\sim 6 \times
10^5\,\msun$. They found that the accretion rate remains below
$\dmcrit = 3 \times 10^{-2}\,\msun\,{\rm yr}^{-1}$ over $1\,{\rm
  Myr}$, leading to their conclusion that accounting for the bloated
state of the VMS is unnecessary. The discrepancy between their
findings and our results primarily arises from the difference in the
initial mass density of the star clusters. While
\cite{2025ApJ...994...40P} consider a central mass density of $\sim
10^7\,\msun\,{\rm pc}^{-3}$, our simulations employ a density of $\sim
10^7\,\msun\,{\rm pc}^{-3}$ within the half-mass radius; consequently,
the central mass density in our models is significantly higher than
$10^7\,\msun\,{\rm pc}^{-3}$. Furthermore, \cite{2025A&A...704A.321V}
identified a VMS exceeding $5 \times 10^4\,\msun$ in a $\sim 6 \times
10^5\,\msun$ star cluster with a half-mass radius density of $\sim 7
\times 10^7\,\msun\,{\rm pc}^{-3}$, which is much higher than our
cluster models. Although they did not incorporate the bloated state,
it is possible that the VMS in their model would satisfy the criteria
for such a state.

\section{Discussions}
\label{sec:Discussions}

It should be noted that the VMS masses and spins obtained in our
simulations represent upper limits, as several mass and angular
momentum loss processes were not explicitly modeled. Stellar
collisions themselves can induce mass loss; therefore, the mass of a
collision product may be less than the combined mass of the two
pre-collision stars \citep{2005MNRAS.358.1133F}. Furthermore, in the
bloated state, the smaller binding energy of the star could
potentially enhance mass loss triggered by mergers
\citep{2006MNRAS.366.1424D, 2009A&A...497..255G, 2023MNRAS.519.5191B,
  2025A&A...699A.223R, 2026A&A...707A.163R}.  Assuming that a VMS
rotates rigidly, we compare the VMS rotation frequency ($\omega$) with
the Kepler rotation frequency ($\omega_{\rm k}$):
\begin{eqnarray}
  \frac{\omega}{\omega_{\rm k}} &=& \frac{J/(km_{\rm VMS}R_{\rm
      VMS}^2)}{\sqrt{Gm_{\rm VMS}/R_{\rm VMS}^3}} \sim
  \sqrt{2}k^{-1}\mave^{1/2}m_{\rm VMS}^{-1/2} \nonumber \\
  &\sim& 1.09 \left( \frac{k}{0.075} \right)^{-1} \left(
  \frac{\mave}{10\;\msun} \right)^{1/2} \left( \frac{m_{\rm VMS}}{3
    \times 10^3\;\msun} \right)^{-1/2}.
\end{eqnarray}
Here, $k$ denotes the moment of inertia coefficient. The second
equality is derived by substituting Eq. (\ref{eq:angularmomentum1})
into the expression for $J$. For the third equality, we adopt $k =
0.075$, which corresponds to a polytrope with index $n = 3$ or,
equivalently, a radiative sphere. Since the VMS rotation frequency
exceeds a few 10 \% of the Kepler rotation frequency, mass shedding is
likely to happen during the evolution of VMSs from hydrogen burning to
helium and carbon burning phases in which the stellar core becomes
compact (see below). A similar reduction likely applies to the spin
angular momentum of the VMSs.

Furthermore, we do not account for stellar winds specifically
associated with VMSs \citep{2018A&A...615A.119V}, nor do we consider
the potentially enhanced stellar winds unique to the bloated state
\citep{2013MNRAS.434.3497G, 2020ApJ...902...81N,
  2021MNRAS.505.2186D}. In the high-luminosity regime typical of VMSs,
accounting for the $\Gamma\mbox{-}\Omega$ limit would presumably
enhance the angular momentum loss as well as the mass loss
\citep{1998A&A...329..551L}.

The dimensionless spin parameter of each VMS is found to be of order
$10$. VMSs evolve from the MS phase to helium and carbon burning ones,
eventually forming an oxygen-carbon core surrounded by an extended
hydrogen and helium envelope~\citep{2018ApJ...857..111T}. The
oxygen-carbon core finally becomes unstable to the electron-positron
pair creation instability if the core mass is smaller than $10^4
M_\odot$~\citep{Shibata:2024xsl}, and if the core mass is large enough
($\agt 140M_\odot$), a BH would be formed. The dimensionless spin for
the oxygen-carbon core is unlikely to be as large as $\sim 10$ because
of the angular momentum transport during the stellar
evolution~\citep{2018ApJ...857..111T}, but even when it is $1$--2, the
remnant after the collapse is likely to be a rapidly spinning BH
surrounded by a massive accretion disk~\citep{2026ApJ...996...57S}.
Such BH-accretion disk systems are capable of driving explosions
fueled by viscous heating and angular momentum transport
\citep{2019ApJ...870...98U}, and may launch jets exhibiting ultra-long
gamma-ray bursts \citep{2011ApJ...726..107S, 2015ApJ...810...64M} if
the jets can penetrate the massive envelope. These could be EM
transients. Furthermore, these massive accretion disks may experience
one-armed spiral instabilities, emitting burst GWs similar to those
associated with GW190521 and GW231123 \citep{2021PhRvD.103f3037S,
  2026ApJ...996...57S}. Consequently, the collapse of a highly
spinning VMS could manifest as a multi-messenger transient.

Note that the VMS reaches a mass of at least $8.4 \times 10^4\,\msun$
in the MS stage for model M6D7H-Y. Such a star should be categorized
as a supermassive star (SMS), which is subject to general-relativistic
radial instability \citep{1963ApJ...138.1090I, 1964ApJ...140..417C,
  1971reas.book.....Z, 1983bhwd.book.....S, 1986ApJ...307..675F}.  The
collapse of an SMS may also be observed as an EM
transient~\citep{Jockel:2025hla}.  Regardless of its spin parameter,
an SMS is believed to undergo a thermonuclear explosion if the
metallicity is high with $Z \agt 0.005$ \citep{1973ApJ...183..941F,
  1986ApJ...307..675F, 2012ApJ...749...37M, 2020MNRAS.496.1224N,
  2022MNRAS.517.1584N, 2023MNRAS.523.1629N, 2024PhRvD.110c1301N} or if
the SMS mass has a particular mass \citep{2014ApJ...790..162C}. If an
SMS is rapidly rotating, a strong shock wave is expected to form
within the accreting material surrounding the BH remnant, driving a
powerful outflow \citep{2006ApJ...641..961L, 2007PhRvD..76h4017L,
  2017PhRvD..96h3016U, 2025ApJ...981..119F}.

We note that it remains non-trivial whether these VMSs/SMSs rotate
rigidly. Since their spin angular momentum is primarily contributed by
stellar collisions, it is possible that only their stellar envelopes
are rapidly rotating, and thus, mass loss due to the stellar wind may
be more serious. To rigorously assess whether these VMSs/SMSs undergo
rigid rotation, three-dimensional hydrodynamics simulations would be
required to investigate the effect of angular momentum
transport. Recent numerical simulations have suggested that collisions
between MS stars amplify the magnetic fields of the resulting products
\citep{2019Natur.574..211S, 2025ApJ...980L..38R,
  2025arXiv251213424O}. Such magnetic fields may facilitate the
transport of angular momentum from the stellar envelope to the
core. In any case, our cluster simulation results highlight the
possibility that VMSs/SMSs formed in dense star clusters could be
eventually observed as EM and GW transients, powered by the resulting
BH-accretion disk systems. Even if they do not explode, they could
potentially become a quasi-star or a BH envelope and grow into
LRDs. At this time, the envelope is expected to rotate
sub-Keplerianly; otherwise, mass shedding would occur.

\section{Conclusions}
\label{sec:Conclusions}

We have investigated the mass and spin growth of VMSs in dense star
clusters using gravitational $N$-body simulations, with parameters
inspired by clusters as potential progenitors that subsequently evolve
into LRDs. Our models incorporate the bloated state of a VMS at the
Hayashi track resulting from successive stellar collisions. We find
that a single VMS typically grows to $\sim 10^3$--$10^4\,\msun$ in
each cluster. Depending on the cluster parameters, the inclusion of
the bloated state enhances VMS growth by several tens of percent to
several times. This demonstrates the necessity of accounting for the
bloated phase when investigating VMS evolution in environments
resembling LRDs.  Furthermore, these VMSs are rapidly rotating, with
dimensionless spin parameters of $\sim 10$ across all investigated
models. This high spin is well-explained by the conversion of the
orbital angular momentum of collided stars into the spin angular
momentum of the VMS. However, it should be noted that our current
models do not account for the mass and angular momentum losses
associated with mergers \citep{2005MNRAS.358.1133F,
  2006MNRAS.366.1424D, 2009A&A...497..255G, 2023MNRAS.519.5191B,
  2025A&A...699A.223R, 2026A&A...707A.163R}, as well as various types
of stellar winds relevant to VMSs \citep{1998A&A...329..551L,
  2013MNRAS.434.3497G, 2018A&A...615A.119V, 2020ApJ...902...81N,
  2021MNRAS.505.2186D}. Detailed modeling of these processes will be
required in future work.

The masses and spins of these VMSs satisfy the necessary conditions
for rotation-driven EM and GW transients, which could be observable
after the collapse of the VMSs.  A highly spinning VMS is expected to
undergo an explosion driven by a BH-accretion disk system formed upon
its collapse. Such systems are also potential sources of burst GWs,
similar to those associated with GW190521 and GW231123. Or possibly,
they could become a quasi-star or a BH envelope and grow into LRDs.

Several physical processes were not explicitly included in this study,
such as mass and angular momentum loss during stellar collisions, the
internal redistribution of spin angular momentum, and post-MS
evolution. Incorporating these effects will be the next crucial step
in refining our understanding of VMS mass and spin evolution.

\begin{acknowledgments}
  We highly appreciate the anonymous referee's valuable
  suggestions. AT is grateful to Takashi Hosokawa for numerous helpful
  advice. This research is supported partly by Grants-in-Aid for
  Scientific Research, 22H00130 (KI), 23H04900 (MS and KI),
  23H05430(KI), 23H01172 (KI), 24K07040 (AT), 25K01035 (AT), and
  26H02045 (KI). AT also thanks for the Step-up program at Fukui
  Prefectural University. Numerical simulations are carried out on the
  Yukawa Institute Computer Facility. We started this study thanks to
  the YITP workshop ``Exploring Extreme Transients 2025''
  (YITP-W-25-08).
\end{acknowledgments}

\begin{contribution}
  A. Tanikawa performed the calculations, analyzed the data, produced
  the figures, and led the manuscript writing. All authors contributed
  to scientific discussion and manuscript writing.
\end{contribution}


\end{CJK*}
\end{document}